\documentclass[referee]{raa} 
\usepackage{graphicx,times}
\usepackage{natbib}
\usepackage{amssymb,amsmath}
\usepackage{overpic}
\usepackage{makecell}
\usepackage{multirow}
\bibpunct{(}{)}{;}{a}{}{,}

\usepackage{hyperref}
\hypersetup{colorlinks = true, linkcolor = green, anchorcolor = red, citecolor = blue, filecolor = red, urlcolor = red}

\begin{document}

\title{Chaos-induced resistivity in different magnetic configurations}
\volnopage{Vol.0 (20xx) No.0, 000--000}      
\setcounter{page}{1}          

\author{Zhen Wang \inst{1,2} \and De-Jin Wu \inst{1} \and Ling Chen \inst{1} \and Yu-Fei Hao \inst{1}}
\institute{Key Laboratory of Planetary Sciences, Purple Mountain Observatory, Chinese Academy of Sciences, Nanjing 210023, China; {\it djwu@pmo.ac.cn}\\
\and University of Science and Technology of China, Hefei 230026, China\\
\vs\no {\small Received~~20xx month day; accepted~~20xx~~month day}}

\abstract{ It is widely believed that magnetic reconnection plays an important role in various eruptive
phenomena of space and astrophysical plasmas. The mechanism of anomalous resistivity,
however, has been an open and unsolved problem. The chaos-induced resistivity proposed
by \cite{Yoshida1998} is one of possible mechanisms for anomalous resistivity. By use of
the test particle simulation, the present work studies the chaos-induced resistivity for different
configurations of reconnection magnetic fields and its distribution in different chaos regions
of reconnection current sheets. The results show that the chaos-induced resistivity can be
$6-7$ orders of magnitude higher than the classical Spitzer resistivity in the X-type chaos
regions and $5$ orders of magnitude in the O-type chaos regions. Moreover, in the X-type
chaos regions the chaos-induced resistivity of the magnetized case is higher by a factor of
2 to 3 times than that of the unmagnetized case, but in the O-type chaos regions the chaos-induced
resistivity of the magnetized case is close to or lower than that of the unmagnetized case.
The present work is helpful to the understanding of the dynamics of reconnection current
sheets, especially of the generation mechanism of the anomalous resistivity of collisionless
reconnection regions.
}

\authorrunning{Z. Wang, D.-J. Wu, L. Chen and Y.-F. Hao}   
\titlerunning{Chaos-induced resistivity and magnetic configurations}

\maketitle

\section{Introduction}
\label{sect:intro}

The early concept of magnetic reconnection was proposed by Giovanelli in 1946 to explain
energy release in solar flares (\citealt{Giovanelli1946}). Nowadays it has been extensively
believed that the magnetic reconnection is one of the most effective processes of magnetic
energy releasing in various eruptive phenomena (\citealt{Ma2018, Lu2018}), such as terrestrial
aurora (\citealt{Dungey1961, McPherron1979}), solar flares (\citealt{Parker1957}), $\gamma$-ray
bursts (\citealt{Lyutikov2006}) and the instability in fusion devices (\citealt{Furth1963}). The
magnetic reconnection process reconfigures the magnetic field topology and converts the
magnetic energy into the kinetic energy of the plasma particles via the heating or acceleration
(\citealt{Dungey1953}). This indicates the breaking of the frozen-in condition, which requires
an enough high plasma resistivity much larger than the classical Coulomb collision resistivity
(i.e., so-called the Spitzer resistivity) in the cases of solar flares and terrestrial aurora (\citealt{Spitzer1956, Cai1997}).
Therefore a collisionless resistivity, usually called the anomalous resistivity or effective
resistivity, is necessary for the collisionless magnetic reconnection (\citealt{Ma2018}).

Many different mechanisms have been proposed to be responsible for the generation the
anomalous resistivity. For example, \cite{Huba1980} pointed out that lower-hybrid drift
waves driven by the diamagnetic current can possibly contribute to the generation of the
anomalous resistivity in an inhomogeneous plasma, \cite{Bulanov1992}  showed that the
electron magnetohydrodynamics effect may lead to the formation of the anomalous resistivity,
and \cite{Biskamp1995} proposed that the Hall effect caused due to the finite ion inertia
also is one of possible mechanisms for the anomalous resistivity. In addition, the presence
of the MHD turbulence, which consists of stochastic magnetic fields over a wide scale range
in the MHD inertial-scale region, may cause an ``effective'' reconnection of the stochastic
magnetic fields, called the turbulent reconnection (\citealt{str88,laz99,laz20}). In this scenario,
however, the so-called turbulent reconnection does not lead to the real dissipation of the
energy in the inertial region, but merely the transport of the energy from larger to smaller
scales due to the turbulent cascade process in the inertial region, and ultimately to the kinetic
scales (\citealt{kim01}).

Some investigations of laboratory experiments and kinetic simulations also showed that
in the kinetic scales the small-scale turbulent magnetic fields typically have two-dimensional
current sheet or three-dimensional flux rope structures (\citealt{gek11,liu13}). The real dissipation
does occur at the kinetic scales of particles in a plasma, at which the motion of individual
particles plays an important role. The kinetic mechanism of producing the anomalous resistivity
that has been investigated most widely is the wave-particle interaction of turbulent plasma
waves in the kinetic scales, such as ion-acoustic turbulence (\citealt{Bychenkov1988}), kinetic
Alfv\'en waves (\citealt{Voitenko1995,Singh2007}), lower-hybrid waves (\citealt{Carter2002}),
and whistler waves (\citealt{Deng2001}).

In fact, the physical nature of the anomalous resistivity is the randomization of the directional
motion of the current-carrying charged particles in the reconnection current sheet. It had
been recognized for a long time that in the neighborhood of the magnetic neutral point of
a spatially inhomogeneous magnetic field, the orbital motion of charged particles possibly
becomes a chaotic motion due to the strong gradient nonlinearity (\citealt{Grad1962, Schmidt1962}).
\cite{Yoshida1998} pointed out that the chaotic motion of charged particles may be equivalent
to the randomization (or i.e., thermalization) process of the directional motion of the current-carrying
particles and hence also can contribute to the generation mechanism of the anomalous resistivity
in the case of collisionless magnetic reconnection. \cite{Numata2002, Numata2003} further
investigated the anomalous resistivity caused by the chaotic motion of particles in an open
system with the convection of particles into and out of the reconnection region. Their results
showed that the continuous dissipation can be carried out and the resulting chaos-induced
resistivity may reasonably explain the necessary anomalous resistivity leading to a fast magnetic
reconnection.

However, in many realistic space and astrophysical plasma environments there frequently
exists a mean magnetic field at a given scale, instead of a three-dimensional magnetic neutral
point (i.e., a real magnetic null point). This indicates that the major magnetic reconnections
are only partly magnetic reconnections with a nonzero guide magnetic field and the so-called
magnetic neutral point actually is a two-dimensional X-type magnetic neutral point in the
reconnection plane perpendicular to the guide field. \cite{Shang2017} generalized the works
of \cite{Numata2002,Numata2003} to the case with a guide field and found that the presence
of a guide field can significantly influences the chaos-induced resistivity nearby the X-type
magnetic neutral point and leads to the chaos-induced resistivity reaching its peak value
when the guide field approaches half of the characteristic strength of the reconnection
magnetic field.

Of course, actually configurations of reconnection magnetic fields in space and astrophysical
plasmas usually are very complex and diverse and the most common one of them is small-scale
current sheets, which may be produced by local plasma currents due to ubiquitous turbulent
magnetic fields (\citealt{str88,laz99,gek11,liu13,laz20}). In this paper we extend the previous
works (\citealt{Shang2017}) to the case of reconnection current sheets, in which there is a finite
width magnetic neutral current sheet between two Y-type magnetic neutral points. In particular,
the distribution of the chaos-induced resistivity in the reconnection current sheet is investigated
and compared to the case of the magnetic configuration with a single X-type magnetic neutral
point. The results show that the chaos-induced resistivity tends to concentrate in the neighborhood
of the X- or Y-type neutral points, where the motion of particles gets into a chaotic orbit more
easily. The possibility of application to the anomalous resistivity of magnetic reconnection in
solar flaring plasmas is discussed. The present results are helpful to our understanding of the
magnetic reconnection physics, especially to the generation mechanism of the anomalous resistivity
in the collisionless magnetic reconnection.

The remains of this paper are organized as follows. The basic physics model and method
used in this investigation are described in section 2, then the results and discussion are
presented in section 3, and finally section 4 devotes to the summary and conclusion.

\section{Basic physics model and method} 
\label{sect:model}

The main aim of this work is to compare the chaos-induced resistivity in different
magnetic configurations of reconnection fields. For the sake of convenience, we use
a magnetic field model with three Cartesian rectangular components, $\boldsymbol{B}=\left(B_x,~B_y,~B_z\right)$,
as follows:
\begin{equation}
\boldsymbol{B}=
\begin{cases}
\left[\tanh\left(y/R_0\right),~\tanh\left(x/R_0-l/2\right),~\delta\right]B_0, & x \ge \left(l/2\right)R_0\\
\left[\tanh\left(y/R_0\right),~\tanh\left(x/R_0+l/2\right),~\delta\right]B_0, & x \leq -\left(l/2\right)R_0\\
\left[\tanh\left(y/R_0\right),~0,~\delta\right]B_0, &  \left| x \right| < \left(l/2\right)R_0
\end{cases},
\end{equation}
where the components in $x-y$ plane,  $B_x$ and $B_y$, is the reconnection field,
the component along the $z$ axis, $B_z$, is the uniform guide field, $B_0$ and $R_0$
are the characteristic strength and the characteristic gradient scale of the reconnection
magnetic field, respectively, $\delta$ is the relative strength of the guide field in units
of $B_0$, and $l$ is the relative width of the neutral current sheet in units of $R_0$.
When $l=0$, the reconnection field has the X-type magnetic configuration and when
$l\to\infty$, the reconnection field becomes one dimensional Harris current sheet.
For a finite $0<l<\infty$, the reconnection field describes a neutral current sheet
with a finite width of $lR_0$ between the two Y-type magnetic neutral points, called
the double Y-type magnetic configuration.

In addition, a constant acceleration electric field along the guide magnetic field
is used, which is given by
\begin{equation}
\boldsymbol{E}
=\left(0,~0,~M_A\right)v_AB_0,
\end{equation}
where $M_A$ is the relative strength of the acceleration electric field $E_z$ in units
of $v_AB_0$ and $v_A$ is the Alfv\'en velocity. Thus, in the collisionless condition
(i.e., to neglect the interactions among particles), the motion equations of individual
charged particles can be described as
\begin{equation}
\begin{aligned}
{\mathrm{d}{\boldsymbol{r}}\over\mathrm{d}t} &=\boldsymbol{v},\\
{\mathrm{d}{\boldsymbol{v}}\over\mathrm{d}t} &=\frac{q}{m}(\boldsymbol{E}+\boldsymbol{v}\times\boldsymbol{B}),
\end{aligned}
\end{equation}
where $\boldsymbol{B}$ and $\boldsymbol{E}$ are the magnetic and electric filed
in Eqs. (1) and (2), respectively.

For convenience used in the numerical simulation, we introduce normalized
variables as follows:
\begin{equation}
\begin{aligned}
& \boldsymbol{r}'=\boldsymbol{r}/R_0,~t'=t/\tau_A,~\tau_A=R_0/v_A \\
& \boldsymbol{v}'=\boldsymbol{v}/v_A,~E_z'=E_z/v_AB_0=M_A,~\boldsymbol{B}'=\boldsymbol{B}/B_0,
\end{aligned}
\end{equation}
where the variables with the superscript ``$'$'' are the dimensionless forms
of the corresponding variables. Thus, the dimensionless forms of the electric
and magnetic fields given by Eqs. (2) and (1) may be written as follows:
\begin{equation}
\boldsymbol{E}'=\left(0,~0,~M_A\right)
\end{equation}
and
\begin{equation}
\boldsymbol{B}'=
\begin{cases}
\left[\tanh\left(y'\right),~\tanh\left(x'-l/2\right),~\delta\right], & x' \ge l/2\\
\left[\tanh\left(y'\right),~\tanh\left(x'+l/2\right),~\delta\right], & x' \leq -l/2\\
\left[\tanh\left(y'\right),~0,~\delta\right], &  \left| x' \right| < l/2
\end{cases},
\end{equation}
respectively. In particular, the dimensionless form of the motion Eq. (3)
can reduce to
\begin{equation}
\begin{aligned}
{\mathrm{d}\boldsymbol{r}'\over\mathrm{d}t'} &=\boldsymbol{v}', \\
{\mathrm{d}\boldsymbol{v}'\over\mathrm{d}t'} &=\omega_c\tau_A\left(M_A+\boldsymbol{v}'\times\boldsymbol{B}'\right).
\end{aligned}
\end{equation}
In order to emphasize the effect of dynamical chaos of particle orbits, we take
$R_0=\lambda_c=v_A/\omega_c$, which implies $\omega_c\tau_A=1$.

Following \cite{Numata2003}, \cite{Andriyas2014},
and \cite{Shang2017}, the anomalous resistivity due to the chaotic motion of
particles may be calculated from the ``average'' acceleration of particles along the acceleration
electric field, $\alpha$, and the relative escape rate of particles from the chaos region, $\beta$,
which are defined by following expressions
\begin{equation}
\bar{v}'_z(t')=\alpha t'
\end{equation}
and
\begin{equation}
n(t')=n_0\exp\left(-\beta t' \right),
\end{equation}
respectively, where $\bar{v}'_z(t')$ and $n(t')$ are the ``average'' velocity and number
density of accelerated particles in the chaos region, respectively. For a given chaos region,
a steady equilibrium can be sustained by continuously supplementing the chaos region
with new particles with zero average velocity. Thus the total average velocity of all particles
in the chaos region may be given by
\begin{equation}
\bar{V}'_z=\frac{\alpha}{\beta}\left[1-\exp\left(-\beta t' \right) \right].
\end{equation}
In particular, a steady average velocity $\bar{V}'_s\equiv\bar{V}'_z\vert_{t'\to\infty}=\alpha/\beta$
can be reached when $t'\to\infty$. On the other hand, in the steady equilibrium the Ohm's
law leads to
\begin{equation}
\bar{V}'_s=\frac{M_A}{\nu_{\rm eff}'}\Rightarrow \nu_{\rm eff}'=\frac{\beta}{\alpha}{M_A},
\end{equation}
where $\nu_{\rm eff}'=\nu_{\rm eff}/\omega_c$ is the dimensionless effective collision
frequency. The corresponding chaos-induced resistivity $\eta_{\rm eff}$ is
\begin{equation}
\eta_{\rm eff}={m\over n_0e^2}\nu_{\rm eff}={B_0\over en_0}\nu_{\rm eff}'.
\end{equation}

In simulations, $2\times 10^5$ particles are initially selected and distributed uniformly
in the region ($-1.0<x',~y'<1.0,~z'=0$) for the X-type magnetic configuration and in the
region ($-5.0<x'<5.0$, $-1.0<y'<1.0$, $z'=0$) for the double Y-type magnetic configuration.
These particles have an initially isotropic Maxwell velocity distribution, i.e.,
\begin{equation}
f(v)=\left({1\over\sqrt{\pi}v_T}\right)^3\exp\left(-{v^2\over v_T^2}\right),
\end{equation}
where $v_T = 0.3v_A$ is a thermal velocity. By using the Runge-Kutta method with an
adaptive time step (the initial step $\delta t'=0.01$) (\citealt{Press1996}), we can trace
the particles in the chaos region based on the motion Eq. (7) and calculate their ``average''
acceleration $\alpha$ in Eq. (8) and the relative escape rate $\beta$ in Eq. (9). Then, the
corresponding effective collision frequency $\nu_{\rm eff}'$ and the chaos-induced resistivity
$\eta_{\rm eff}$ can be obtained from Eqs. (11) and (12), respectively.

\section{Results and discussion} 
\label{sect:results}
\subsection{The Case of in X-type Magnetic Configuration with $l=0$} 

Figure 1 shows the X-type magnetic field configuration, where $R_c$ is the radius of chaos
region (i.e., the shadow area in the $x-y$ plane) in the unit of $R_0=\lambda_c$. The guide
filed $B_z$ and the the acceleration electric field $E_z$ are normalized by the reconnection
field $B_0$ and the Alfv\'enic induced field $v_AB_0$, respectively. For the given chaos region
radius $R_c/R_0=1.0$ and three fixed values of the acceleration electric field $E_z'=M_A=0.0001$,
0.0005, and 0.001, the ``average'' acceleration $\alpha$, the relative escape rate $\beta$,
and the effective collision frequency $\nu_{\rm eff}'$ versus the guide field $B_z$ from 0
to $2.0B_0$ are displayed in panels (a), (b), and (c) of Figure 2, respectively, where the
parameters $\alpha$ and $\beta$ are obtained by means of fitting the average velocity
along the z axis and the number of the particles surviving inside the chaos region based
on Eqs. (8) and (9), respectively, and $\nu_{\rm eff}'$ is obtained from Eq. (11).

\begin{figure}[htbp]
\centering
\includegraphics[width=6.5cm]{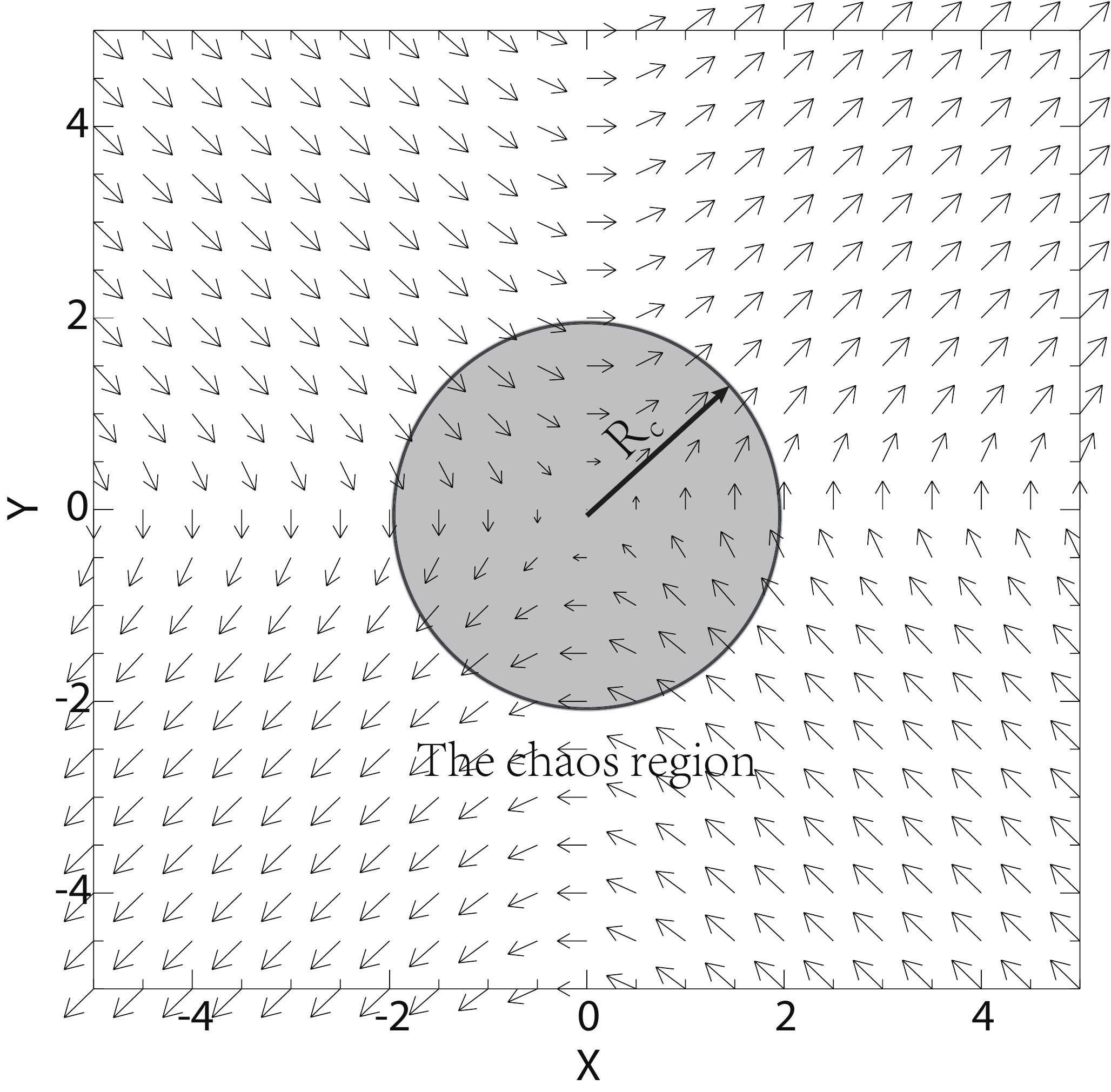}
\caption{The X-type magnetic configuration and the chaos region (i.e., shadow area),
where $R_c$ is the radius of the chaos region.}
\end{figure}

From Fig. 2 (a), the ``average'' acceleration $\alpha$ increases with the acceleration electric
field (i.e., $M_A$), implying that the acceleration of electrons is dominated by the acceleration
electric field in the presence of the guide field, as expected by previous works (\citealt{Wan2008,
Huang2010, Wang2017, Lu2018, Xia2018}). Fig. 2 (b) shows that the relative escape rate
$\beta$ also increases with the parameter $M_A$ because the larger acceleration electric
field possibly leads to particles escaping more quickly from the chaos region. For the lower
guide field case of $B_z/B_0<1$, the relative escape rate $\beta$ decreases considerably
as the guide field increases, as shown in Fig. 2 (b). This is probably because the guide field
can effectively prevent particles from escaping from the given chaos region. However, the
effect of the guide field on the particle escaping is significantly weakened when the guide
field is stronger than the reconnection field, that is, $B_z>B_0$. Fig. 2 (c) gives the effective
collision frequency $\nu_{\rm eff}'$, which has been calculated by Eq. (11). As expected
by the proportional relation in Eq. (11), the parameter $\nu_{\rm eff}'$ in Fig. 2 (c) has
similar changes to the parameter $\beta$ in Fig. 2 (b).

\begin{figure*}
\begin{center}
\begin{tabular}{ccc}
\begin{overpic}[width=.45\textwidth]{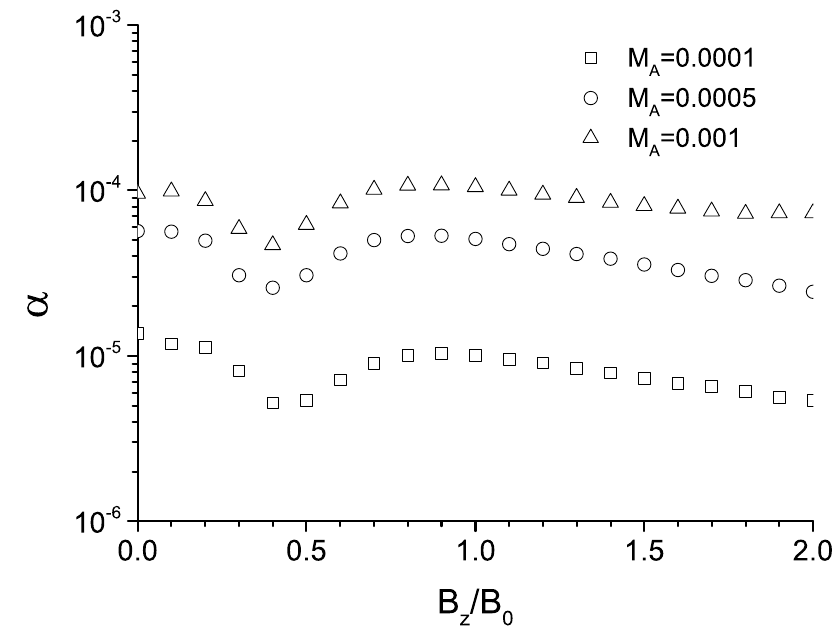}\put(50,-5){(a)}\end{overpic}\\[6pt]
\begin{overpic}[width=.45\textwidth]{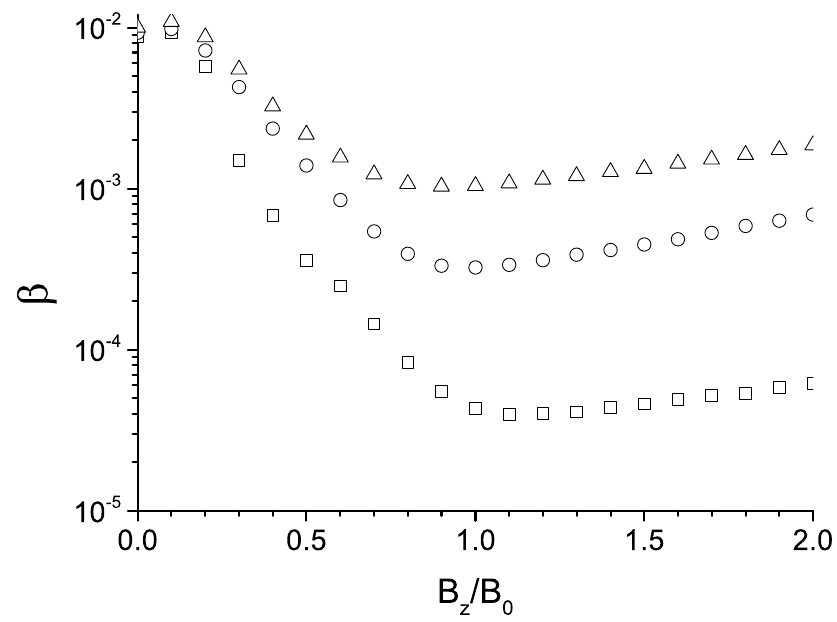}\put(50,-5){(b)}\end{overpic}\\[6pt]
\begin{overpic}[width=.45\textwidth]{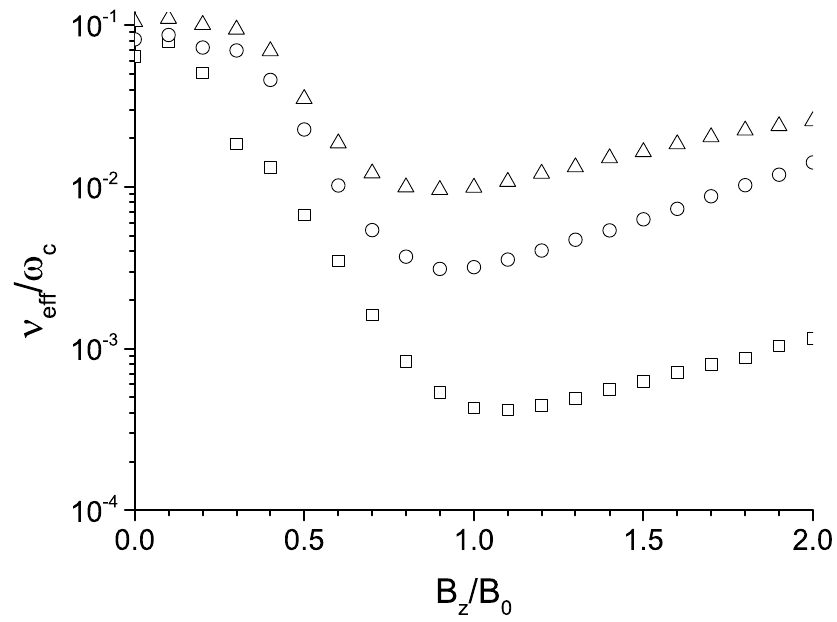}\put(50,-5){(c)}\end{overpic}
\end{tabular}
\caption{The ``average'' acceleration $\alpha$, the relative escape rate $\beta$, and the
                   effective collision frequency $\nu_{\rm eff}'$ versus the guide field $B_z$ in
                   different acceleration electric fields with $M_{\rm A}=0.0001, 0.0005$, and
                   $0.001$.}
\end{center}
\end{figure*}

Figure 3 displays the same parameters as Fig. 2 but for the fixed guide field $B_z=0.5B_0$
and varying the chaos region radius $R_c$ from $0.1$ to $2.0R_0$. From Fig. 3, it can be found
that the parametric variation in smaller chaos regions of $R_c<R_0$ are considerably different
from that in larger regions of $R_c>R_0$. For the smaller regions of $R_c<R_0$, the ``average''
acceleration $\alpha$ increases with the region radius $R_c$, the relative escape rate $\beta$
is nearly a constant, and hence the effective collision frequency $\nu_{\rm eff}'$ decreases
with the chaos region radius $R_c$. For the larger regions of $R_c>R_0$, on the other hand,
the ``average'' acceleration $\alpha$ and the relative escape rate $\beta$ both decreases with
the region radius $R_c$, but the effective collision frequency $\nu_{\rm eff}'$ increases with
the chaos region radius $R_c$.

\begin{figure*}
\begin{center}
\begin{tabular}{ccc}
\begin{overpic}[width=.45\textwidth]{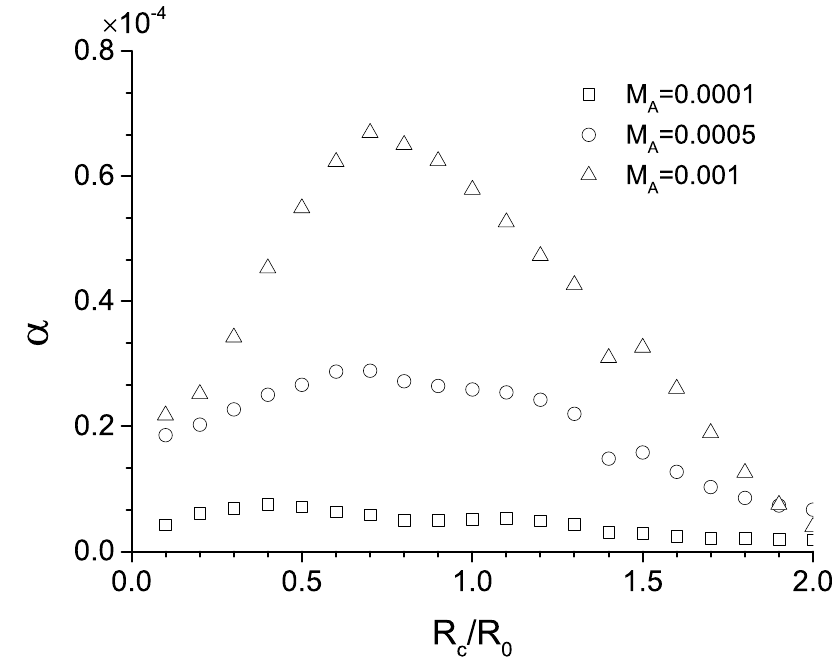}\put(50,-5){(a)}\end{overpic}\\[6pt]
\begin{overpic}[width=.45\textwidth]{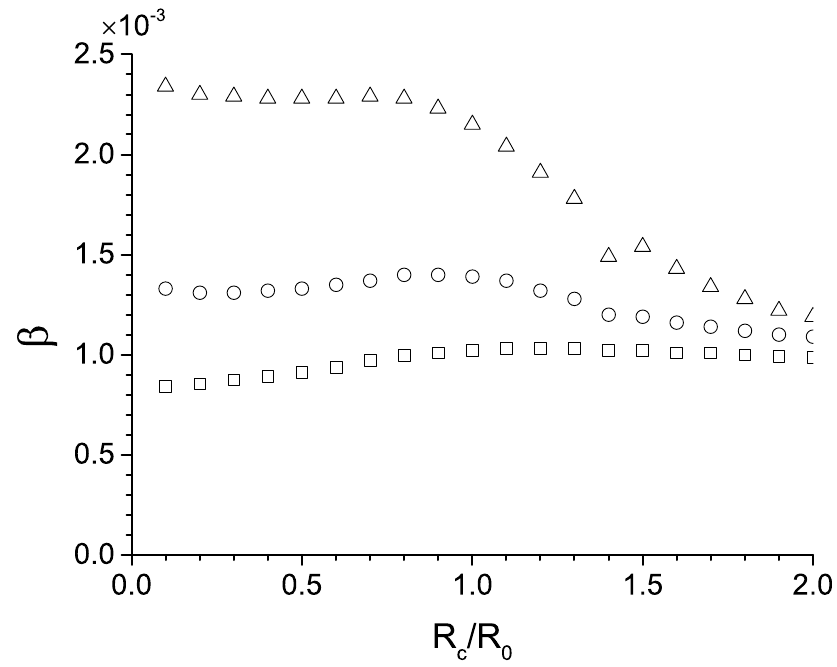}\put(50,-5){(b)}\end{overpic}\\[6pt]
\begin{overpic}[width=.45\textwidth]{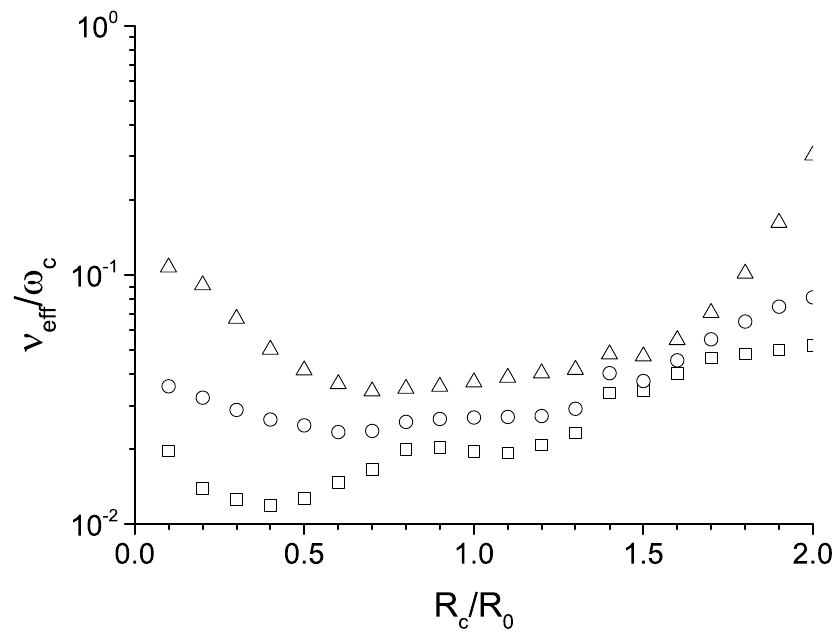}\put(50,-5){(c)}\end{overpic}
\end{tabular}
\caption{The ``average'' acceleration $\alpha$, the relative escape rate $\beta$, and the
                   effective collision frequency $\nu_{\rm eff}'$ versus the guide field $B_z$ in
                   different acceleration electric fields with $M_{\rm A}=0.0001, 0.0005$, and
                   $0.001$.}
\end{center}
\end{figure*}

For the parameters used in the present simulation, we have $R_0=v_A/\omega_c=\left(v_A/v_T\right)r_L\simeq 3.3 r_L$,
where $r_L=v_T/\omega_c$ is the cyclotron radius of particles. The Larmor cyclotron effect
of particle motion due to the Lorentz force influence generally the acceleration and escape
processes of the particles in the larger magnetized regions of $R_c>R_0$, which size is considerably
larger than the particle cyclotron radius. However, the Larmor cyclotron motion of particles
has little effect on their acceleration and escape in the smaller unmagnetized region of $R_c<R_0$
because the size of their moving region is less than the cyclotron radius. In the unmagnetized
region of $R_c<R_0$, the ``average'' acceleration $\alpha$ increases with $R_c$ because the particles
have longer effective acceleration times for the larger chaos region. While in the magnetized
region of $R_c>R_0$ the acceleration and escape of more particles are hindered by the cyclotron
motion for the more chaos region so that the ``average'' acceleration $\alpha$ and the escape
rate $\beta$ decrease with $R_c$.

On the other hand, the distribution of the effective collision frequency $\nu_{\rm eff}'$
in Fig. 3(c) indicates that the chaos-induced resistivity has an inhomogeneous distribution
in the reconnection region, in which the chaos-induced resistivity is higher inside the
unmagnetized region of $R_c<R_0$ and outside the magnetized of $R_c>R_0$, but lower
in the transition region of $R_c\sim R_0$. In addition, the variation of the effective collision
frequency $\nu_{\rm eff}'$ with the acceleration electric field, as shown in Fig. 2(c) and
Fig. 3(c), implies the nonlinear characteristic of the Ohm's law associated with the chaos-induced
resistivity, which may be attributed to the influence of the acceleration electric field on
the chaos dynamics of the particles and hence on the chaos-induced resistivity in the reconnection
region.

\subsection{The Case of in Double Y-type Magnetic Configuration with $l=8$} 

The double Y-type magnetic configuration is shown in Figure 4, which includes two Y-type
magnetic neutral points at $x\pm l/2$ and $y=0$ and a magnetic neutral line at $y=0$,
and the later denotes a current sheet with a width $l=8$. The three typical chaos regions
are displayed by the shadow areas in the panels (a), (b), and (c) of Fig. 4, respectively,
in which the shadow area in the panel (a) displays the chaos regions around the Y-type
neutral points (at $x\pm l/2$, $y=0$), the shadow area in the panel (c) does the typical
chaos region in the neutral current sheet (at $x=0$, $y=0$), and the shadow area in the
panel (b) contains the whole reconnection current sheet.

\begin{figure*}
\begin{center}
\begin{tabular}{ccc}
\begin{overpic}[width=.45\textwidth]{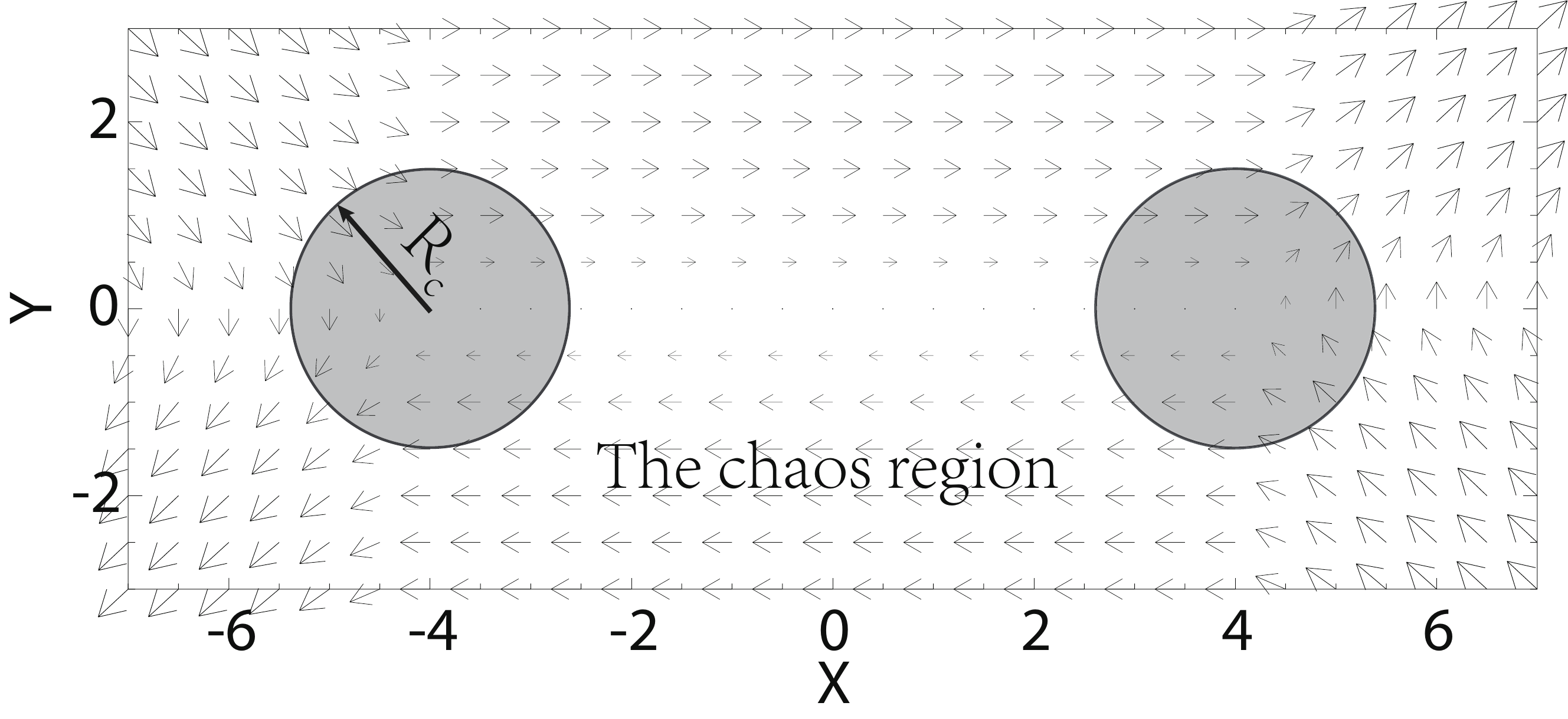}\put(50,-5){(a)}\end{overpic}\\[8pt]
\begin{overpic}[width=.45\textwidth]{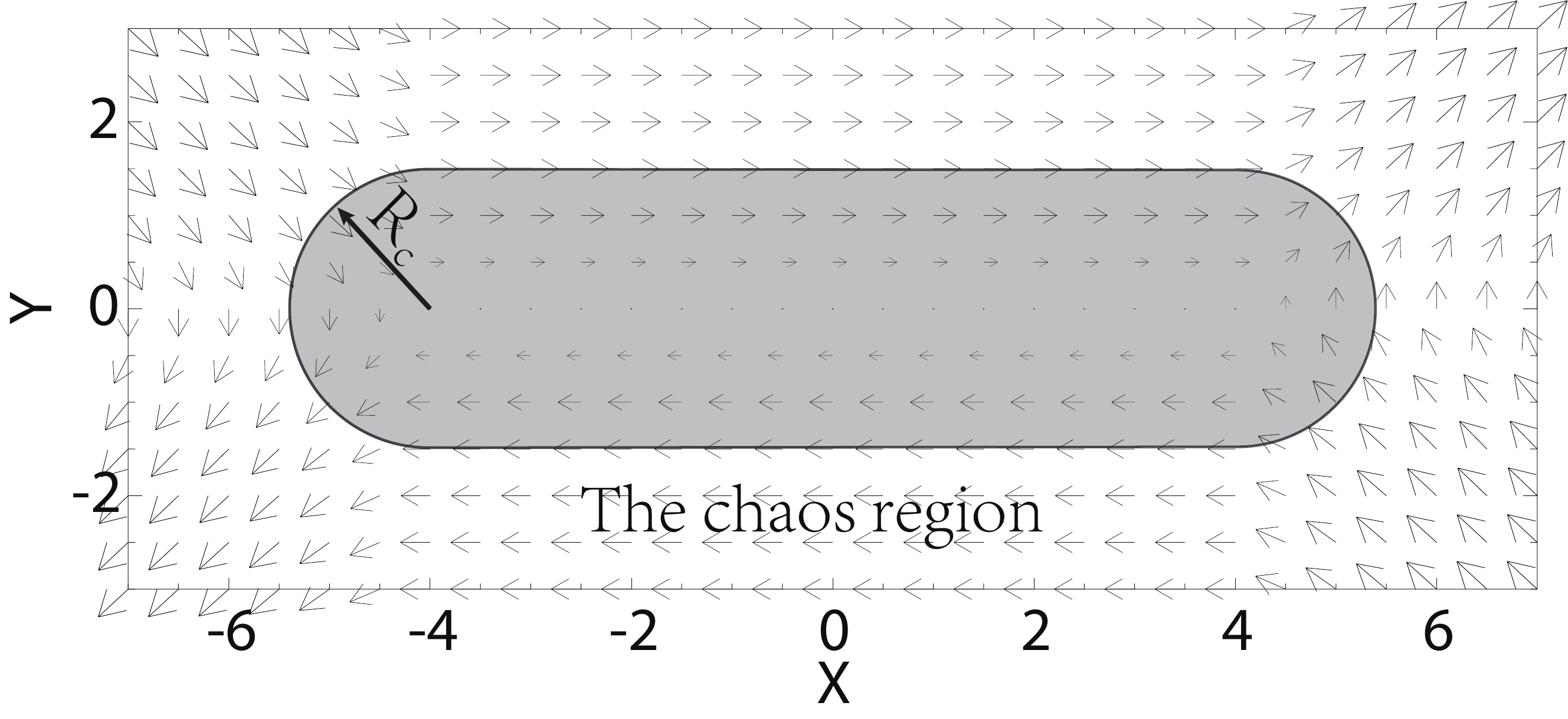}\put(50,-5){(b)}\end{overpic}\\[8pt]
\begin{overpic}[width=.45\textwidth]{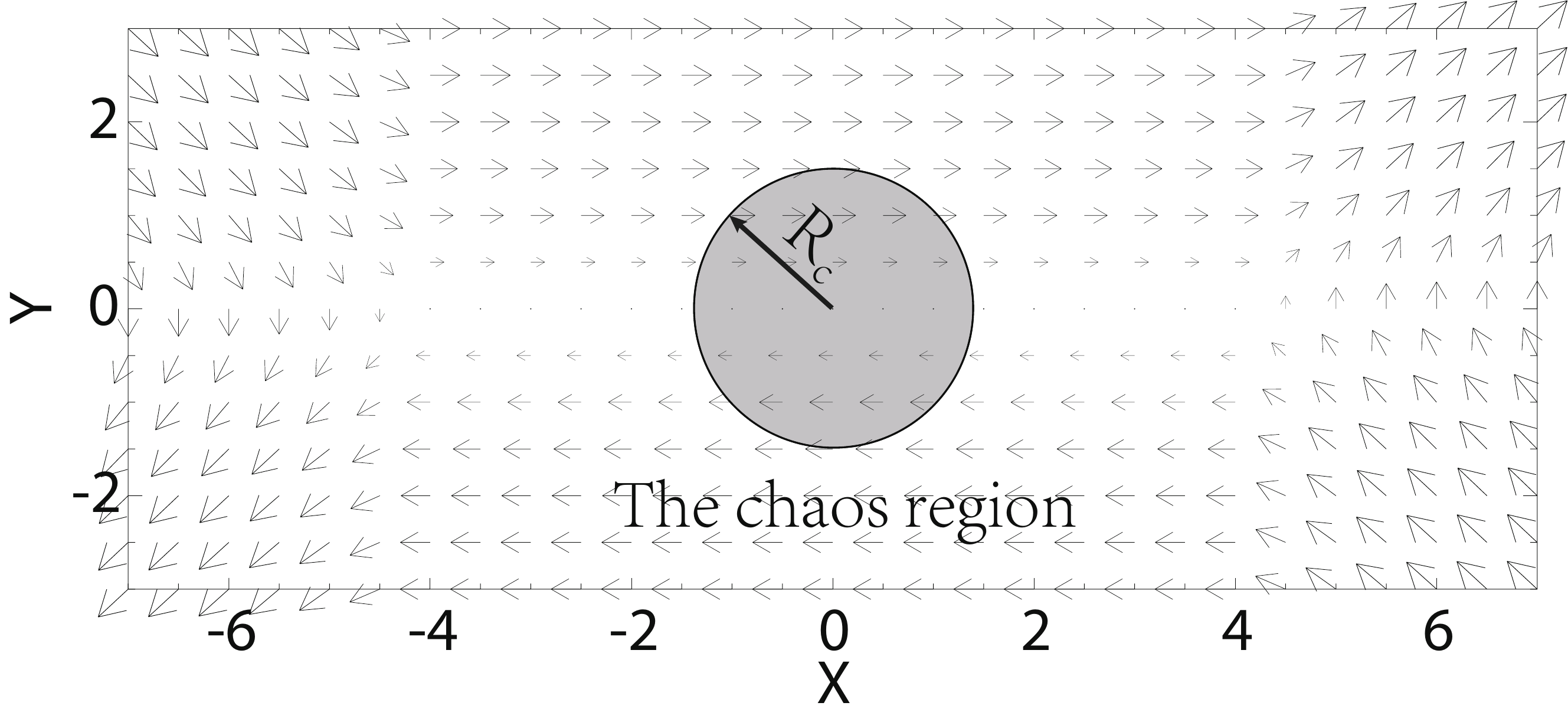}\put(50,-5){(c)}\end{overpic}
\end{tabular}
\caption{The ``average'' acceleration $\alpha$, the relative escape rate $\beta$, and the
                   effective collision frequency $\nu_{\rm eff}'$ versus the guide field $B_z$ in
                   different acceleration electric fields with $M_{\rm A}=0.0001, 0.0005$, and
                   $0.001$.}
\end{center}
\end{figure*}

Figure 5 plots the ``average'' acceleration $\alpha$, the relative escape rate $\beta$, and
the effective collision frequency $\nu_{\rm eff}'$ versus the guide field $B_z$ for three
fixed values of the acceleration electric field $M_A=0.0001$, 0.0005, and 0.001 and the
given chaos region radius $R_c=1.0R_0$, in which the left, middle, and right columns are
corresponding to the chaos regions in the panel (a), (b), and (c) of Fig. 4, respectively.
It is similar to the case of the X-type magnetic configuration presented in Fig. 2 that the
``average'' acceleration $\alpha$, the relative escape rate $\beta$, and the effective collision
frequency $\nu_{\rm eff}'$ all increase with the acceleration electric field $M_A$. Moreover,
for the stronger acceleration electric field of $M_A=0.001$, $\alpha$ increases considerably
with the guide field $B_z$, which is different from that of the X-type magnetic configuration
and further confirms that the presence of the guide field will be favorable to the particle
acceleration by the acceleration electric field in the reconnection current sheet (\citealt{Wan2008,
Huang2010, Wang2017, Lu2018, Xia2018}).

In particular, it is worthy to notice that the ``average'' acceleration $\alpha$ and the relative
escape rate $\beta$ increases and decreases, respectively, in turn from left to right columns.
As expected by Eq. (11), this directly results in the effective collision frequency $\nu_{\rm eff}'$
(or i.e., the corresponding chaos-induced resistivity) in the neutral current sheet chaos region
in Fig. 4(c) significantly lowering than that in the Y-type chaos regions shown in Fig. 4(a) and
much less than that in the X-type chaos region presented in Fig. 1 and Fig. 2(c). One of possible
reasons is that the particle motion near the X-type neutral point getting into a chaotic orbit
more easily (\citealt{Shang2017}).

\begin{figure}[htbp]
\centering
\includegraphics[width=16cm]{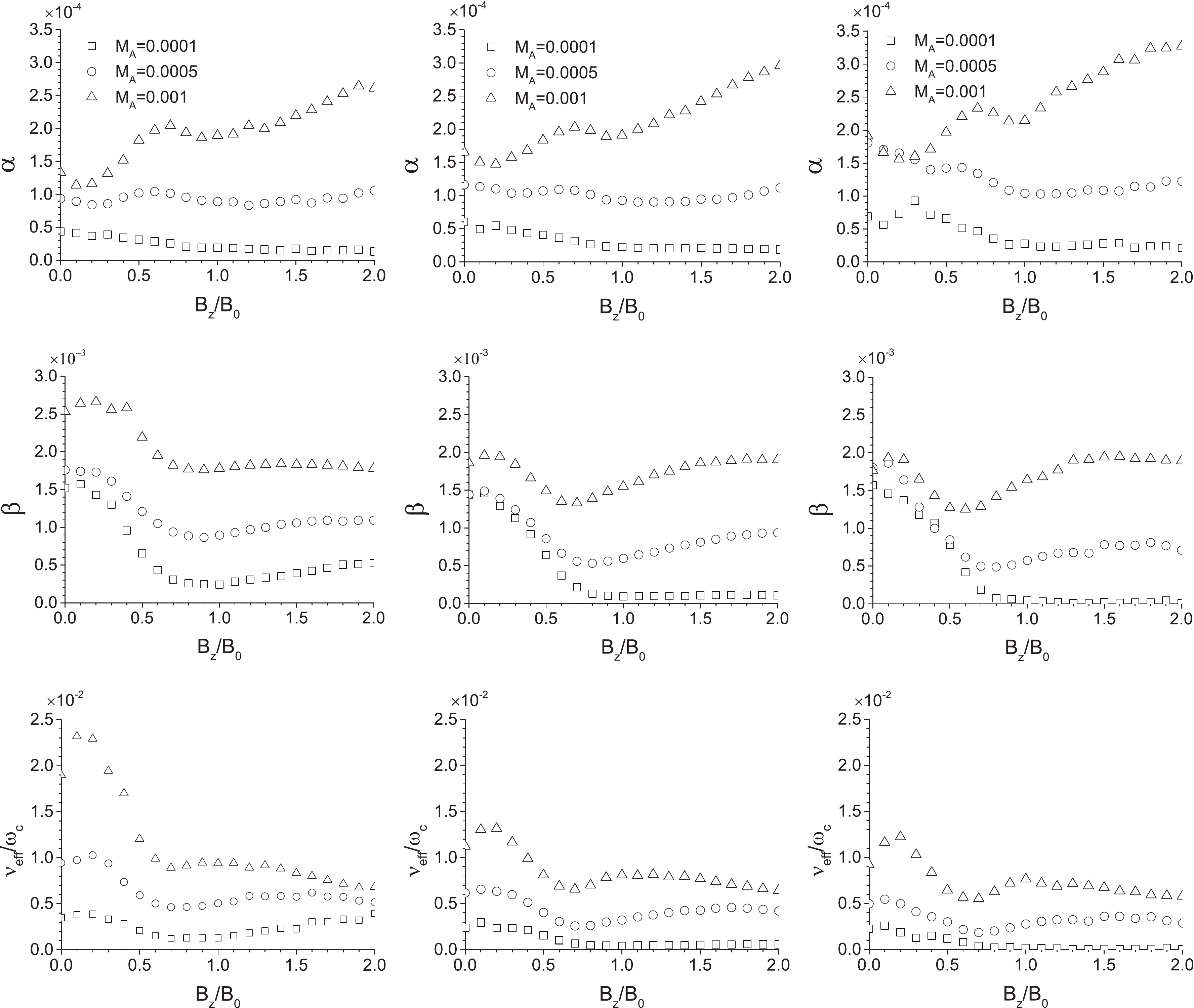}
\caption{The ``average'' acceleration $\alpha$, the relative escape rate $\beta$, and the effective
collision frequency $\nu_{\rm eff}'$ versus the guide field $B_z$ for the given chaos region
radius $R_c=1.0R_0$ and three fixed acceleration electric fields $M_A=0.0001$, 0.0005, and
0.001. The left, middle and right columns present the results of the chaos  regions (a), (b),
and (c) in Fig. 4, respectively.}
\end{figure}

\begin{figure}[htbp]
\centering
\includegraphics[width=16cm]{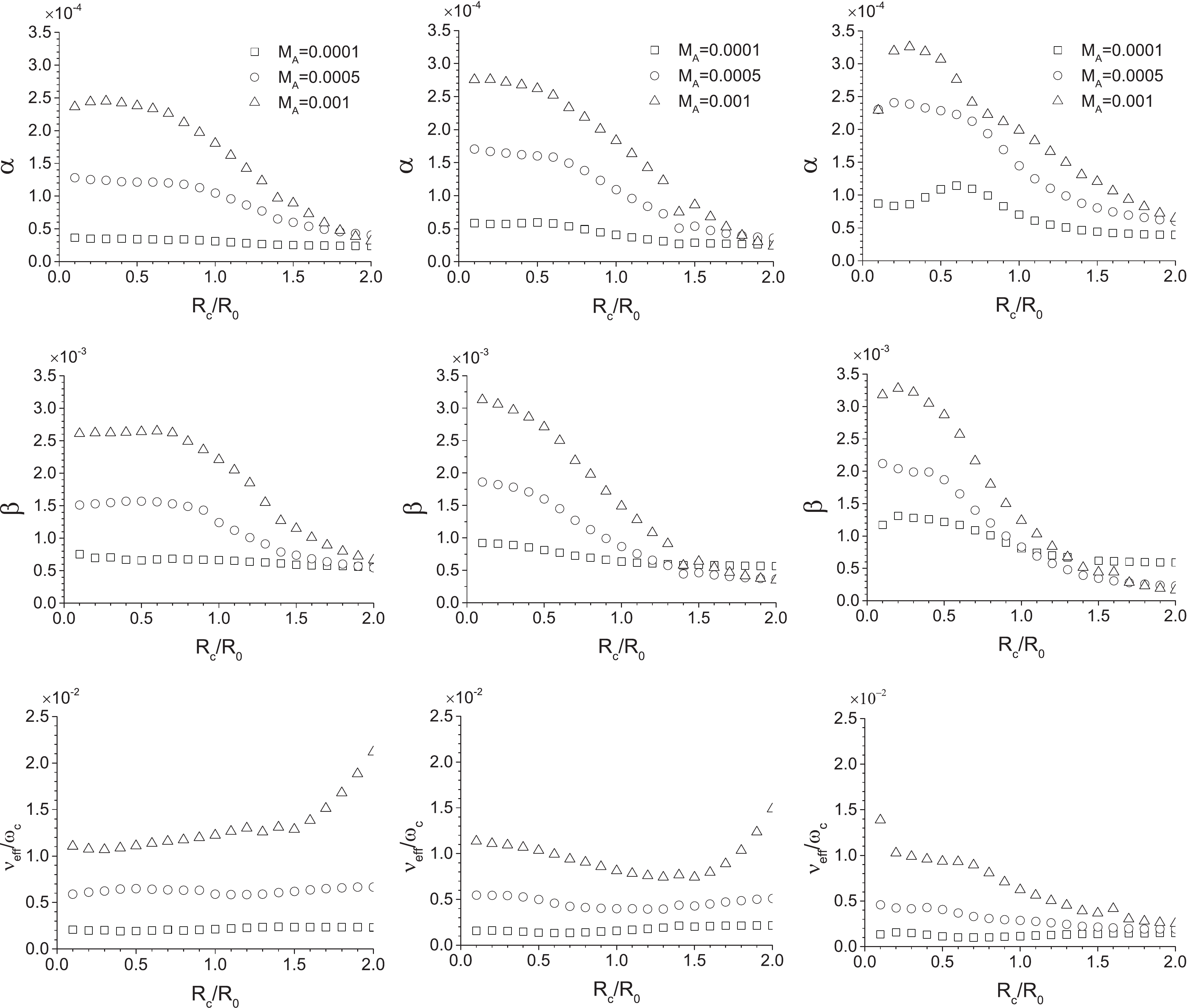}
\caption{The ``average'' acceleration $\alpha$, the relative escape rate $\beta$, and the effective
collision frequency $\nu_{\rm eff}'$ versus the chaos region radius $R_c$ for the given guide
field $B_z=0.5B_0$  and three fixed acceleration electric fields $M_A=0.0001$, 0.0005, and 0.001.
The left, middle and right columns present the results of the chaos  regions (a), (b), and (c)
in Fig. 4, respectively.}
\end{figure}

For the given guide field $B_z=0.5B_0$, Figure 6 presents the ``average'' acceleration $\alpha$,
the relative escape rate $\beta$, and the effective collision frequency $\nu_{\rm eff}'$ versus
the chaos region radius $R_c$ for the three fixed values of the acceleration electric fields
$M_A=0.0001$, 0.0005, and 0.001, in which the left, middle and right columns are corresponding
to the calculation results from the chaos  regions (a), (b), and (c) in Fig. 4, respectively. From
Fig. 6, it can be found that the ``average'' acceleration $\alpha$ and relative escape rate $\beta$
are determined mainly by the acceleration electric field $M_A$ and are little influenced by the
radius $R_c$ in the unmagnetized region of $R_c<R_0$. In the magnetized region of $R_c>R_0$,
however, $\alpha$ and $\beta$ both decrease considerably as the radius $R_c$ increases
because the cyclotron motion due to the reconnection field can depress the acceleration and
escape of the particles.

Meanwhile, Fig. 6 shows that the ``average'' acceleration $\alpha$ in the central chaos region
of the neutral current sheet (in the right column) is higher than that in the Y-type chaos
regions (in the left column), similar to Fig. 5. However, different from Fig. 5, the relative
escape rate $\beta$ in the central chaos region is higher than that in the Y-type chaos regions
for the unmagnetized case ($R_c<R_0$) but lower for the magnetized case ($R_c>R_0$).
This implies that the particles escape more easily from the unmagnetized region along the
parallel reconnection field lines in the current sheet, but more difficultly from the magnetized
region due to the hindrance of the cyclotron motion. Moreover, also Fig. 6 shows that the
effective collision frequency $\nu_{\rm eff}'$, and hence the chaos-induced resistivity, in
the Y-type chaos regions (on the left column) is higher considerably than that in the central
chaos region of the reconnection current sheet (on the right column), especially for the magnetized
region of $R_c>R_0$ and the stronger acceleration electric field $M_A$.

In addition, it is also worthy to notice that the  effective collision frequency $\nu_{\rm eff}'$
increases and decreases with the chaos region radius $R_c$ in the  Y-type chaos region and
in the central chaos region, respectively. It has a potential importance of understanding the
dynamics of reconnection current sheets that there are considerably different chaos-induced
resistivity in the different types of chaos regions in reconnection current sheets. A lot of
investigations have shown that an initial reconnection current sheet with the double Y-type
magnetic configuration usually is unstable against the resistive tearing mode, especially in
high-temperature plasmas such as the solar corona (\citealt{Sato1979, Ugai1984, Schumacher1996,
Nitta2007, Ma2015, Ma2018, Wang2017, Lu2018, Xia2018}). This may lead to the formation
of a magnetic island, which has an O-type neutral point at its center and two X-type neutral
points at its two sides, between the two Y-type neutral points, and then the magnetic island
further collapses to a new current sheet with the double Y-type configuration. In consequence,
the development of this instability possibly leads to the occurrence of a island chain with
alternate O-type and X-type neutral points. In general, the particle motion near the
O-type and X-type neutral points can have significantly different characteristics \citep{Fu2006}.

In the present simplified model, the magnetic island and O-type neutral point can be represented
by the central chaos region (i.e., shadow area in Fig. 4(c)) and the central point (at $x=0$, $y=0$)
in the reconnection current sheet, respectively. The evolution processes of the reconnection
current sheet mentioned above, in general, are accompanied by the locally enhancements of
the current density at the O-type neutral points and of the dissipation at the X-type neutral
points, respectively (\citealt{Schumacher1996}).  While the locally enhancements of the current
density at the O-type neutral points and of the dissipation at the X-type neutral points indicate
that the reduction and the rise of the chaos-induced resistivity (i.e. the anomalous resistivity)
in the O-type chaos regions and the X-type chaos regions, respectively. Also this is consistent
with the expected results of the effective collision frequency shown in Figs. 3 and 6 (see further
discussions of the chaos-induced resistivity in the next subsection).

\subsection{Possible Application in Solar Coronal Plasmas} 

It is believed extensively that solar flares are produced by magnetic field reconnections in
the solar corona. From Eqs. (11) and (12), the effective collision frequency due to the effect
of the particle orbit chaos can be calculated by
\begin{equation}
\nu_{\rm eff}=\nu_{\rm eff}'\omega_c={eB_0\over m}\nu_{\rm eff}',
\end{equation}
and the corresponding chaos-induced resistivity may be given by
\begin{equation}
 \eta_{\rm eff}={m\over n_0e^2}\nu_{\rm eff}={B_0\over en_0}\nu_{\rm eff}'.
\end{equation}
For typical parameters of pre-flaring coronal plasmas, such as plasma density $n_0=10^{16}$
m$^{-3}$, temperature $T=500$ eV, and magnetic field $B_0=100$ G, Table 1 shows
the weighted average of the effective collision frequencies over the chaos region radius
$R_c$ from $R_c=0.1R_0$ to $R_c=1.0R_0$ for the unmagnetized case,
\begin{equation}
\bar{\nu}_{\rm eff}'\equiv\sum_{r_c=0.1}^{1.0}{r_c^2\nu_{\rm eff}^\prime (r_c)}\left(\sum_{r_c=0.1}^{1.0}{r_c^2}\right)^{-1},
\end{equation}
and from $R_c=1.1R_0$ to $R_c=2.0R_0$ for the magnetized case,
\begin{equation}
\bar{\nu}_{\rm eff}'\equiv\sum_{r_c=1.1}^{2.0}{r_c^2\nu_{\rm eff}^\prime (r_c)}\left(\sum_{r_c=1.1}^{2.0}{r_c^2}\right)^{-1},
\end{equation}
where $r_c\equiv R_c/R_0$, $r_c^2$ is proportional to the number of particles inside the
radius $R_c$ and represents the weight of the region with the radius $R_c$. For the sake
of comparison, Table 1 also presents the ratio of  the corresponding average chaos-induced
resistivity to the classical Spitzer resistivity due to the Coulomb collision, $\eta_{\rm eff}/\eta_{\rm cal}$,
where the average chaos-induced resistivity is given by $\eta_{\rm eff}=\left(B_0/en_0\right)\bar{\nu}_{\rm eff}'$
and the classical Spitzer resistivity $\eta_{\rm cal}$ for the above typical parameters of
pre-flaring coronal plasmas can be estimated approximately as $4.65\times 10^{-8}~({\rm\Omega\cdot m})$
by the Spitzer formula (\citealt{Spitzer1956}) with the Coulomb logarithm $\ln\Lambda\approx 10$.

\begin{table}
\centering
\caption[]{The average effective collision frequency $\bar{\nu}_{\rm eff}'$ and the radio of the
                  chaos-induced to the classical Spitzer resistivity $\eta_{\rm eff}/\eta_{\rm cal}$ for
                  given acceleration electric field $M_A=0.0001$, 0.0005, and 0.001 in different chaos
                  regions, where ``X-type'', ``Y-type'', ``S-type'', and ``O-type'' stand for the chaos regions
                  shaded in Fig. 1, Fig. 4(a),  Fig. 4(b), and Fig. 4(c), respectively, and the typical parameters
                  of pre-flaring coronal plasmas have been used in the calculation of the classical Spitzer resistivity.}\label{Tab:publ-works}
\begin{center}
\begin{tabular}{|p{0.8cm}<{\centering}|p{1.8cm}<{\centering}|p{1.3cm}<{\centering}|p{1.3cm}<{\centering}|p{1.3cm}<{\centering}
|p{1.7cm}<{\centering}|p{1.7cm}<{\centering}|p{1.7cm}<{\centering}|}
\hline
\multicolumn{2}{|c|}{} & \multicolumn{3}{|c|} {$\bar{\nu}_{\rm eff}'$} & \multicolumn{3}{|c|}{$\eta_{\rm eff}/\eta_{\rm cal}$} \\
\hline
\multicolumn{2}{|c|}{$M_A$} & 0.0001 & 0.0005 & 0.001 & 0.0001 & 0.0005 & 0.001 \\
\hline
\multirow{2}{*}{X-type} & unmagnetized & 0.01789 & 0.02585 & 0.03835 & $2.41\times10^6$ & $3.47\times10^6$ & $5.15\times10^6$ \\
\cline{2-8}
 & magnetized & 0.04105 & 0.05504 & 0.11460 & $5.51\times10^6$ & $7.39\times10^6$ & $1.54\times10^7$ \\
\hline
\multirow{2}{*}{Y-type} & unmagnetized & 0.00205 & 0.00624 & 0.01175 & $2.75\times10^5$ & $8.38\times10^5$ & $1.58\times10^6$ \\
\cline{2-8}
 & magnetized & 0.00234 & 0.00637 & 0.01594 & $3.14\times10^5$ & $8.55\times10^5$ & $2.14\times10^6$ \\
\hline
\multirow{2}{*}{S-type} & unmagnetized & 0.00146 & 0.00427 & 0.00906 & $1.96\times10^5$ & $5.73\times10^5$ & $1.22\times10^6$ \\
\cline{2-8}
 & magnetized & 0.00206 & 0.00461 & 0.00999 & $2.77\times10^5$ & $6.19\times10^5$ & $1.34\times10^6$ \\
\hline
\multirow{2}{*}{O-type}& unmagnetized & 0.00110 & 0.00324 & 0.00784 & $1.47\times10^5$ & $4.35\times10^5$ & $1.05\times10^6$ \\
\cline{2-8}
 & magnetized & 0.00143 & 0.00211 & 0.00345 & $1.92\times10^5$ & $2.83\times10^5$ & $4.63\times10^5$ \\
\hline               
\end{tabular}
\end{center}
\end{table}

In Table 1, the four types of chaos regions shown in Fig. 1, Fig. 4(a),  Fig. 4(b), and Fig. 4(c)
are denoted by ``X-type'', ``Y-type'', ``S-type'', and ``O-type'', respectively, and each chaos
region is divided into unmagnetized ($R_c<R_0$) and magnetized ($R_c>R_0$) cases, where
the three acceleration electric fields $M_A=0.0001$, 0.0005, and 0.001 have been used too.
From Table 1, one can find that in a high-temperature plasma such as the pre-flare coronal
plasmas, the chaos-induced resistivity can produce an effective resistivity much higher than
the classic collisional resistivity by $5-7$ orders of magnitude, implying that the chaos-induced
indeed may provide an efficient mechanism for the anomalous resistivity in collisionless
reconnection regions (\citealt{Shang2017}).

However, as shown in \S 3.2, the chaos-induced resistivity can have remarkable different
values in the different chaos regions of the reconnection current sheet. For a high-temperature
plasma such as the pre-flare coronal plasma, the results of Table 1 shows that the chaos-induced
resistivity in the X-type chaos region, in general, is significantly higher than that in the O-type
chaos region by one to two orders of magnitude. For instance, from Table 1 it can be found
that for the magnetized case with a stronger acceleration electric field $M_A=0.001$, the ratio
of the chaos-induced resistivity to the classical Spitzer resistivity, $\eta_{\rm eff}/\eta_{\rm cal}$,
in the X-type chaos region is $\sim 1.54\times 10^7$ remarkably larger than $\sim 4.63\times 10^5$
in the O-type chaos region by a factor $\sim 33$. This inhomogeneity of the chaos-induced
resistivity in spatial distribution in reconnection current sheets may be attributed to that
the particle motion near the X-type neutral point, in general, gets into the chaotic orbits more
easily than that near the O-type neutral point (\citealt{Shang2017}).

In fact, some investigations of dynamic current sheets (\citealt{Schumacher1996}) also suggested
that dynamical processes of reconnection current sheets have strongly local dependence, in
which the enhanced current channels are formed mainly near the O-type neutral points and
the major energy dissipation occurs around the X-type neutral points. While the current enhancement
near the O-type neutral point usually implies that there is a lower anomalous resistivity in the
O-type chaos region and the dissipation enhancement around the X-type neutral point generally
implies that there is a higher anomalous resistivity in the X-type chaos region. Therefore, the
present results proposes that the chaos-induced resistivity can be responsible for the generation
mechanism of the anomalous resistivity that well meets the requirements of the dynamics of
the reconnection current sheets.

The inhomogeneity of the chaos-induced resistivity in the reconnection current sheet manifests
not only in the difference among various chaos regions but also in the distinction between the
magnetized and unmagnetized cases in the same chaos region. From Table 1, for the same chaos
region, especially the X-type chaos region with higher chaos-induced resistivity, the chaos-induced
resistivity of the magnetized case is considerably higher than that of the unmagnetized case,
except for the two cases of the O-type chaos region with stronger acceleration electric field
of $M_A=0.0005$ and $0.001$. This indicates that the cyclotron motion of charged particles
in the magnetized case also may contribute to the increase of the chaos-induced resistivity.

In addition, it should be pointed out that the acceleration electric field also can significantly
influence the chaos-induced resistivity. The generation and role of the acceleration electric
field in the dynamical evolution of reconnection current sheets has been a complex problem
(\citealt{Wang2017, Lu2018}). In general, the electric field in the reconnection current sheet
can be created locally by magnetic induced process (i.e., $\nabla\times{\bf E}=-\partial {\bf B}/\partial t$),
such as in the X-type chaos region, or by flow driven process (i.e., ${\bf E}=-{\bf v}\times{\bf B}$),
such as in the O-type chaos region. This indicates that there also can be an inhomogeneous
electric field in the reconnection current sheet. In particular, the electric field, in general, is
greatly dynamical evolution dependent on the dynamical proceeding of the reconnection
current sheet. In fact, the present results also show that the chaos-induced resistivity can
have considerable variation when the acceleration electric field varying because the electric
field may greatly influence the chaotic motion of charged particles. Therefore, the electric
field is possibly an important uncertain factor in the dynamics of the reconnection current
sheet and more further investigations are needed. Moreover, not all acceleration effects can
be included by an equivalent ``acceleration electric field'' although the electric field possibly
plays the most important role in the acceleration (\citealt{Wang2017, Lu2018}).

\section{Summary and conclusion}
\label{sect:summary}

The anomalous resistivity has been an open problem in the collisionless magnetic reconnection,
which is commonly believed to be responsible for the magnetic energy releasing mechanism
of solar flares. Investigations on the dynamics of reconnection current sheets (\citealt{Sato1979,
Ugai1984, Schumacher1996, Nitta2007, Ma2015, Ma2018, Wang2017, Lu2018, Xia2018})
showed that the dynamical process has remarkably localized and inhomogeneous characteristics,
in which the enhancements of current and dissipation occurs, respectively, near the O-type
and the X-type neutral points in the reconnection current sheet. In particular, this implies
that the anomalous resistivity should have an inhomogeneous distribution in the reconnection
current sheet, and hence that a reasonable generation mechanism for the anomalous resistivity
should could lead to higher and lower anomalous resistivity near the O-type and X-type neutral
points, respectively.

In the present work, the chaos-induced resistivity has been investigated for different magnetic
configurations, which may be corresponding to different regions of reconnection current sheet.
The present results show that for the case of high-temperature plasmas, such as in the pre-flare
coronal plasmas, the chaos-induced resistivity can be much higher than the classic collisional 
resistivity by a factor $\sim 10^5-10^7$. This indicates it to be possible that the chaos-induced
resistivity provides an enough high anomalous resistivity for the fast magnetic energy release
in solar flares. In particular, the dependence of the chaos-induced resistivity on the magnetic
configuration reveals that the chaos-induced resistivity has a greatly inhomogeneous distribution
in reconnection current sheets, in which, as expected by the evolution of dynamical current
sheets (\citealt{Schumacher1996}), the chaos-induced resistivity in the X-type chaos region
is considerably higher than that in the O-type chaos region of the dynamical current sheet
because the particle motion in the X-type chaos region gets into the chaotic orbits more easily
than that in the O-type chaos region (\citealt{Shang2017}). Therefore, the present results proposes
that the chaos-induced resistivity can be one of possible candidates for the anomalous resistivity
that may meet the requirements of the dynamics of the reconnection current sheets.

As shown in the last section, however, the electric field in the reconnection current sheet
can significantly influence the chaos-induced resistivity because the acceleration of particles
by the electric field probably can directly affect the chaotic motion of the particles. While
usually it is very difficult to determine the distribution of the electric field in the current
sheet. In fact, the electric field, the current, and the chaos-induced resistivity can interact
each other in the dynamics of the reconnection current sheet, and hence the relationship
between them is no longer the linear dependent relation in the common Ohm's law but
nonlinear coupling mutually. Therefore, some more self-consistent study is necessary for
better understanding of the role of the chaos-induced resistivity in the dynamics of reconnection
current sheets. The results by the present simple model proposed that the chaos-induced
resistivity may become one of promising mechanisms for the anomalous resistivity in collisionless
magnetic reconnection phenomena and deserves to pay more attention and further investigation.

\begin{acknowledgements}
This work was supported by National Natural Science Foundation of China (Grant No.
41531071, No. 11873018, No. 11790302 and No. 11761131007). The authors would
like to gratitude to Meng Shang and Xiao-Long Wang for their helps in the simulation.
\end{acknowledgements}

\label{lastpage}

\end{document}